\documentstyle[12pt]{article}

\begin{document}

\topmargin 0pt
\oddsidemargin 5mm
\renewcommand{\thefootnote}{\fnsymbol{footnote}}

\begin{titlepage}
\setcounter{page}{0}

\vspace{1cm}
\begin{center}
{\Large Pseudoclassical theories of Majorana, Weyl\\
and Majorana--Weyl particles}
\vspace{0.5cm}

{\large Grigoryan G.V.\raisebox{.8ex}{$\star$}, Grigoryan R.P.
\raisebox{.8ex}{$\star\star$}, Tyutin
I.V.\raisebox{.8ex}{\ddag}
\footnote{ P.N.Lebedev Physical Institute, Moscow, Russia}}
\footnote{ Partially supported  by the grant 211-5291 YPI
 of the German Bundesministerium f\"ur Forschung und
 Technologie.}\\
\vspace{1cm}

{\em Yerevan Physics Institute, Republic of Armenia}\\

\vspace{1cm}
\raisebox{.8ex}{$\star$}{E-mail:GAGRI@VXC.YERPHY.AM}\\
\raisebox{.8ex}{$\star\star$}{E-mail:ROGRI@VXC.YERPHY.AM}\\
\raisebox{.8ex}{\ddag}\vspace{1cm}
{E-mail:TYUTIN@LPI.AC.RU}
\vspace{5mm}
\end{center}
\centerline{{\bf{Abstract}}}
A pseudoclassical theories of Majorana, Weyl and Majorana--Weyl
particles in the space--time dimensions $D=2n$ are constructed. The
canonical quantization of these  theories is carried out and as
a result we obtain the quantum mechanical description 
of neutral particle in $D=2n$ ,
Weyl particle in $D=2n$ and  neutral Weyl particle in $D=4n+2$.
In $D=2,4({\rm mod}8)$ dimensional space--time the description of the
neutral particle coincides with the field theoretical description
of the Majorana particle in the Foldy--Wouthuysen representation.
In $D=8k+2$ dimensions the neutral Weyl particle coincides with
 the Majorana--Weyl particle in the Foldy--Wouthuysen
representation.

\vfill
\centerline{\large Yerevan Physics Institute}
\centerline{\large Yerevan 1995}

\end{titlepage}
\newpage
\renewcommand{\thefootnote}{\arabic{footnote}}
\setcounter{footnote}{0}

\section{Introduction}
\indent
In spite of the bulk of the papers devoted to the theories of point
particles and to methods of their quantization, the problem still
attracts the attention of the investigators and numerous
classical models of particles and superparticles were discussed
recently. The renewal of the interest to these theories is
primarily due to the problems in the string theory since particles
are prototypes of the strings.

The first pseudoclassical description of the relativistic
spinning particle was given in paper \cite{BM1,BM2}. It was
followed by a great number of papers  [3-22]
 devoted to the different
quantization schemes of that theory, to the introduction of the
internal symmetries and to the generalization to higher spins.

A pseudoclassical theory of Weyl particle in the space--time
$D=4$ was constructed in \cite{GGT}. Their method was used in
\cite{GG8} to  construct  the theory of Weyl particle in the space--time
dimensions $D=2n$, where a quantum mechanics of the
neutral Weyl particle in even--dimensional space--time was
suggested and the connection of this theory  with the theory of
Majorana--Weyl particle in QFT  for $D=10$ was  discussed.

In this paper the description  of Majorana, Weyl and
 Majorana--Weyl particles in the pseudoclassical approach is
 investigated in arbitrary even dimensions. This is interesting
 in connection with the following.

As it is known, the theory of RNS string with a GSO projection
(see. \cite{GSW}) is a supersymmetric theory in ten dimensional
space--time. The supersymmetry requires that each mass level
comprises a supermultiplete. In the massless sector the
superpartners to the gauge vector fields , which in $D=10$
dimensions have eight degrees of freedom, are
massless fermions - Majorana--Weyl bispinors.  Hence it is
interesting to construct the classical (pseudoclassical) model,
which after quantization will bring to the theory of the
Majorana--Weyl bispinor in $D=10$ dimensions.

In section 2 the pseudoclassical theory of $D=2n$ dimensional
 neutral spinning
particle is presented. The canonical quantization of that theory
in arbitrary even dimensional space--time is carried out and it's
found, that for $D=2,4({\rm mod}8)$ the quantum mechanical
description coincides with the field theoretical description of
the Majorana particle in the Foldy--Wouthuysen representation.
In section 3 the pseudoclassical theory of Weyl particle in $D=2n$
dimensional   space--time   is  investigated.  The  canonical
quantization  of  that theory
results  in  the  theory which coincides with field
theoretical   description   of the  $D=2n$  dimensional  Weyl
particle in the Foldy--Wouthuysen representation. The section
4  is  devoted  to the canonical quantization of the neutral
Weyl  particle  in the space--time dimensions $D=4n+2$ and it
is  shown  that  in  the  dimensions  $D=8k+2$  this  theory
coincides with the theory of the Majorana--Weyl particle in the
Foldy--Wouthuysen representation in the field theory.

\section{Even dimensional neutral (Majorana) particle}

The  quantum  mechanical  description  of  Majorana particle
repeats  in  essence  the  similar  description of the Dirac
particle.  The  canonical  quantization of the corresponding
pseudoclassical  theory  of the Dirac particle
\cite{BM1}  was  carried  out  in
\cite{GT2}  in the dimensions $D=4$ and in \cite{GG1} in the
arbitrary  even  dimensions  $D=2n$.  We will write down the
main  important  formulae,  which  will  be used below ( for
details  see  \cite{GT2,GG1}).  The  action of the theory is
given by the expression
\begin{equation}
\label{MA}
S=\frac{1}{2}
\int d\tau \left[\frac{1}{e}\left(\dot{x}^\mu -\frac{i}{2}\chi
\xi^\mu\right)^2+e
m^2+im\chi\xi_{D+1}-i\left(\xi_\mu\dot{\xi}^\mu
-\xi_{D+1}\dot{\xi}_{D+1}
\right)\right].
\end{equation}
Here   $x^\mu$   are  the   coordinates  of  the  particle,
$\mu=0,1,\dots  ,D-1$;  $\xi^\mu$  are  Grassmann  variables
describing the spin degrees of freedom; $e,\,\chi,\,\xi_{D+1}$
are   additional   fields,   $e$   being   Grassmann   even,
$\chi,\,\xi_{D+1}$--Grassmann  odd
variables;   the  overdote  denotes  a  differentiation  with
respect to the parameter $\tau $ along the trajectory.

This  action   is   invariant   under  two  types  of  gauge
transformations:       the       re\-pa\-ra\-me\-tri\-zation
transformations 
\begin{eqnarray}
\label{RT1}
\delta x^\mu&=&u\dot{x}^\mu,  \quad\delta
e=\frac{d}{d\tau}(ue),
\nonumber\\
\delta\xi^\mu&=&u\dot{\xi}^\mu,\quad
\delta\xi_{D+1}=u\dot{\xi}_{D+1},
\quad \delta\chi=\frac{d}{d\tau}(u\chi)
\end{eqnarray}
with    the    even    parameter    $u(\tau)$, and the  supergauge
transformations
\begin{eqnarray}
\label{ST1}
\delta x^\mu&=&iv\xi^\mu,\quad\,\delta e=iv\chi,\quad\,
\quad\delta\chi=2 \dot{v},\nonumber\\
\delta\xi^\mu&=&v\left(\frac{\dot{x}^\mu}{e}-\frac{i\chi}{2e}
\xi^\mu\right),\quad\delta\xi_{D+1}=vm
\end{eqnarray}
with   the   odd   parameter   $v(\tau)$.

The  hamiltonization  of  the theory brings to the canonical
hamiltonian
\begin{equation}
\label{MHAM}
H=\frac{e}{2}\left(p^2-m^2\right)+\frac{i}{2}\chi
\left(p_\mu\xi^\mu-m\xi_{D+1}\right)
\end{equation}
and to the set of primary constraints 
\begin{eqnarray}
\label{PRC}
\Phi_\mu&\equiv&\pi_\mu-\frac{i}{2}\xi_\mu\approx 0,\quad
\mu=0,1,\dots,D-1,\nonumber\\
\Phi_{D}&\equiv&\pi_{D+1}+\frac{i}{2}\xi_{D+1}\approx 0,\quad
\Phi_{D+5}\equiv\pi_e\approx 0,\quad
\Phi_{D+7}\equiv\pi_\chi\approx 0.
\end{eqnarray}
Here $p_\mu,\,\pi_e,\,\pi_\mu,\,\pi_{D+1},\,\pi_\chi$ are
canonical momenta conjugated to
$x^\mu,\,e,\,\xi^\mu,\,\xi_{D+1},\,\chi$ correspondingly. Using
the Dirac  procedure \cite{D1,GTY} we find the secondary constraints
\begin{equation}
\label{SC1}
\Phi_{D+1}\equiv
p_\mu\xi^\mu-m\xi_{D+1}\approx 0,\quad
\Phi_{D+3}\equiv|p_0|-\omega\approx 0;\quad
\omega=\left({p}_{i}^2+m^2\right)^{1/2}.
\end{equation}
The theory contains first class constraints
$\Phi_{D+3},\quad\Phi_{D+5},\quad\Phi_{D+7},\quad$ and
\begin{equation}
\label{FCC1}
 \varphi=p^\mu
\Phi_\mu+m\Phi_D+i\Phi_{D+1}=\frac{1}{2}i\left(p^\mu\xi_\mu
-m\xi_{D+1}\right)+p^\mu\pi_\mu+m\pi_{D+1}.
\end{equation}
The  additional gauge conditions, equal in number to that of
all first class constraints and conjugated to the latter,
which will fix the gauges of the theory, are chosen in the form:
\begin{eqnarray}
\label{GFC}
\Phi_{D+4}&\equiv& x^0-\kappa \tau \approx 0,\quad
\Phi_{D+8}\equiv \chi\approx 0,\quad \Phi_{D+6}\equiv
e+1/|p_0|\approx 0,\nonumber\\
&&\Phi_{D+2}\equiv   \xi^0   \approx   0;\qquad  \kappa=-{\rm
sign}p_0.
\end{eqnarray}
After the canonical transformation
\begin{equation}
\label{CTR}
x^0\rightarrow x^{0\prime}=x^0-\kappa\tau
\end{equation}
(the rest of the variables remain unchanged), bringing to
constraints which do not depend of time explicitly, the theory is
described by the Hamiltonian (primes are omitted)
\begin{equation}
\label{MCH}
H=\omega=\left({p}_{i}^2+m^2\right)^{1/2},
\end{equation}
 and    by   the   independent   variables   $x^i,\,p_i,
\,\xi^i$, for which the  Dirac brackets for complete set of
constraints
have the form:
\begin{eqnarray}
\label{DB}
\{x^i,  x^j\}_D&=&\frac{i}{2\omega^2}[\xi^i,\xi^j]_-,  \quad
\{x^i,  p_j\}_D=\delta^i_j,  \quad\{p_i,  p_j\}_D=0,  \nonumber\\
\{x^i,  \xi^j\}_D&=&\frac{1}{\omega^2}\xi^i p^j ,\quad
\{\xi^i,  \xi^j\}_D=-i\left(\delta^{ij}-\frac{p^i
p^j}{\omega^2}\right),  \quad
\{p_i,  \xi^j\}_D=0.
\end{eqnarray}
The quantization of the theory is carried out through the
realization of the operators
$\hat{x}{}^i,\,\hat{p}{}_i,\,\hat{\xi}{}^i$ in the form:
\begin{eqnarray}
\label{QOR}
\hat{x}{}^i&=&\hat{q}{}^i-\frac{i\hbar}{4}\frac{1}
{\hat{\omega}\left(\hat{\omega}+m\right)}\left[\Sigma^i,\hat{p}_j
\Sigma^j\right],\nonumber\\
\hat{\xi}{}^i&=&\left(\frac{\hbar}{2}\right)^{1/2}\hat{\kappa}
\left[\Sigma^i-\frac{1}
{\hat{\omega}\left(\hat{\omega}+m\right)}\hat{p}_i\hat{p}_j
\Sigma^j\right],\quad \hat{p}_k= -i\frac{\partial}{\partial q^k},
\end{eqnarray}
where operators  $\hat{q}{}^i$ (physical coordinate operators)
are  multiplication  operators,  the  variable  $\kappa$  is
replaced by the operator 
$\hat{\kappa}$
 \begin{equation}
\label{KAP}
\hat{\kappa}=\left(
\begin{array}{cc}
I&0\\0&-I
\end{array}
\right)=\tau^3\otimes I, \quad \hat{\kappa}{}^2=1
\end{equation}
with     eigenvalues $\kappa=\pm 1$,  the   operators    $\Sigma^i={\rm
diag}(\sigma^i,\sigma^i)$,
$\sigma^i$ are $2^\frac{D-2}{2}\times 2^\frac{D-2}{2}$ matrices,
 which realize  the Clifford algebra
 $\left[\sigma^i,\sigma^j\right]_+=2\delta^{ij}I$.

 The wave function $f$ is given by
 \begin{equation}
\label{SV1}
f=\left(
\begin{array}{c}
f^1(q)\\f^2(q)
\end{array}                                      	
\right), 	
\end{equation}
where $f^1$ and $f^2$ are $2^{\frac{D-2}{2}}$ component
columns, and the Shr\"odinger equation has the form
\begin{equation}
\label{SE1}
(i\partial/\partial x^0-\hat{\kappa}\hat{\omega})f=0,\quad
q=(x^0,q^i)
\end{equation}
where we passed from the variable $\tau$ to the
the physical time $x^0=\kappa\tau$.

The canonical generators of the Lorentz transformation 
\begin{equation}
\label{CLG}
J^{\mu\nu}=-\left(x^\mu p^\nu-x^\nu p^\mu
+\frac{i}{2}[\xi^\mu,\xi^\nu]_-\right)
\end{equation}
after  quantization  of the theory, in terms of operators of
the physical variables, are given by the expressions 
\begin{eqnarray}
\label{LG}
\hat{J}^{ik}&=&-\hat{q}^i \hat{p}^k + \hat{q}^k \hat{p}^i -
\frac{i\hbar}{4}\left[\Sigma^i, \Sigma^k\right]_{-},\nonumber\\
\hat{J}^{0k}&=&-x^0 \hat{p}^k
-\frac{1}{2}\hat{\kappa}\left[\hat{q}^k, \hat{\omega}\right]_{+}
-\frac{i\hbar}{4 \hat{\omega}}\hat{\kappa}\hat{p}^j
\left[\Sigma^k, \Sigma^j\right]_- .
\end{eqnarray}
In \cite{GT2,GG1} it is shown, that $f^1$ must be interpreted as
the wave function of the particle, $f^2$ as the complex
conjugated wave function of the antiparticle and the quantum
mechanical description coincides with the description of the
Dirac particle in the field theory in the Foldy--Wouthuysen
representation.

Lets turn now to the construction of the quantum mechanics of the
Majorana particle. Note, that the action (\ref{MA}) is invariant
under the transformations
\begin{eqnarray}
\label{TI}
x^\mu(\tau)&\rightarrow& x^\mu(-\tau),   \quad
\xi^\mu(\tau)\rightarrow \xi^\mu(-\tau),   \quad
\xi_{D+1}(\tau)\rightarrow
-\xi_{D+1}(-\tau),\nonumber\\
\chi(\tau)&\rightarrow&\chi(-\tau),\quad  e(\tau)\rightarrow
e(-\tau),\quad i\rightarrow -i,
\end{eqnarray}
which correspond to the reparametrization $\tau\rightarrow
-\tau$. This transformation was not included in the gauge
group at the beginning of this section. In that case
the model describes the charged particle and in the gauge
$x^0-\kappa\tau\approx 0$ the trajectories with $\kappa=+1$ are
interpreted  as  trajectories  of  particles  and those with
$\kappa=-1$ as trajectories of
antiparticles.  The switching  on of the external electromagnetic
field confirms this assertion since  to the trajectory with a
given $\kappa$ corresponds a particle  with a  charge  $\kappa
e$  \cite{GT2} and the action isn't invariant under the
transformation $\tau\rightarrow -\tau$. When the action is
invariant under the transformation $\tau\rightarrow -\tau$,
there is a possibility of another interpretation. We can
identified the trajectories with $\kappa=+1$ and $\kappa=-1$.
This is equivalent to including of the reparametrization
$\tau\rightarrow -\tau$ in the gauge group \cite{GT2} and then
the theory describes the truly neutral particle.

The quantization of the theory in this case may be carried out in
two ways:

A) to choose a gauge, which violates the
reparametrization symmetry $\tau\rightarrow -\tau$ as well;

     B)  to  quantize  the  theory in the gauge which doesn't
violate   the  reparametrization  symmetry  $\tau\rightarrow
-\tau$  and  then  factorize  with  respect to that symmetry
(i.e.   identify   the  trajectories  with  $\kappa=+1$  and
$\kappa=-1$).

A) In this case the convenient choice of the gauge is the
replacement of the constraint $\Phi_{D+4}$  by the
constraint \begin{equation}
\label{MGF}
\Phi^\prime_{D+4}=x^0-\tau \approx 0.
\end{equation}
The quantum mechanics in this case coincides with the sector
$\kappa=+1$ of the theory described at the beginning of this
section. In particular the wave function $f$ is now a
$2^\frac{D-2}{2}$ component column, which coincides with the
$f^1$:
\begin{equation}
f=f^1.
\end{equation}

For the comparison of this theory with the description of the
neutral (Majorana) particle in the field theory it is convenient
to introduce the "bispinor" $f_M$
\begin{equation}
\label{QMF}
f_M= \left(
\begin{array}{c}
f^1\\\Lambda f^{1*}
\end{array}
\right)
\end{equation}
The choice of the matrix $\Lambda$ will be discussed below.

B)   The   factorization   with   respect  to  the  symmetry
$\tau\rightarrow -\tau$ of the quantum mechanics 
of the Dirac particle consists in fact in
the identification of the sectors $\kappa=+1$ and $\kappa=-1$.
To give the rule of such identification  note, that from the
Shr\"odinger equation (\ref{SE1}) it follows, that $f^1$ contains
only positive frequencies, while $f^2$ contains only negative
ones. Thus the rule must be of the form: $f^2\sim
f^{1*}$. This is in accordance with the interpretation of $f^2$
as a complex conjugated wave function of the antiparticle and
also with the explicit form of the transformation
$\tau\rightarrow -\tau$. Hence the factorization rule means, that
in the state space of the Dirac particle we must restrain
ourselves to the vectors having the form (\ref{QMF}), where
$\Lambda$ is a certain unitary operator, which we'll chose in the
form of the numerical unitary $2^\frac{D-2}{2}\times
2^\frac{D-2}{2}$ matrix. Since $f^1$ and $f^2$ have definite
transformation properties under Lorentz transformations, the
$\Lambda$ matrix must satisfy the relation
\begin{equation}
\label{MLG}
\hat{J}^{\mu\nu}(\kappa=-1)=-\Lambda
\hat{J}^{\mu\nu*}(\kappa=+1)\Lambda^+.
\end{equation}
Using the explicit form (\ref{LG}) of the $J^{\mu\nu}$
generators we find, that $\Lambda$ matrix must have the property
\begin{equation}
\label{LMP}
\left[\sigma^i,\sigma^j\right]_-=\Lambda\left[\sigma^{i*},
\sigma^{j*}\right]_-\Lambda^+,
\end{equation}
whence it follows that
\begin{equation}
\label{SIG}     	
\sigma^i=\epsilon\Lambda\sigma^{i*}\Lambda^+,\quad\epsilon=
+1\quad{\rm or }\quad -1.
\end{equation}
It's not difficult to check, that if such a matrix exists, then
it is unique up to a sign.
We'll give the explicit expressions for the $\Lambda$ matrices
in any even dimensional space--time $D=2n$ (though the $\Lambda$
matrix exists in any dimensions).

We'll choose a special form of the $\Gamma_{(2n)}^{\mu_{(2n)}}$,
$\mu_{(2n)}=0,1,\dots,2n-1$ in the $D=2n$ dimensions in the Dirac
representation, which we'll describe inductively.
Let by definition $\Gamma_{(0)}^{\mu_{(0)}}=1$. The $2\times 2$
dimensional matrices $\Gamma_{(2)}^{\mu_{(2)}} $ are equal to
\begin{equation}
\label{G2}
\Gamma_{(2)}^{0}=\tau^3,\quad \Gamma_{(2)}^{1}=i\tau^2,
\end{equation}
where $\tau^k$, $k=1,2,3$ are Pauli matrices. If the matrices
$\Gamma_{(2n)}^{\mu_{(2n)}}$ are known, then the
$\Gamma_{(2n+2)}^{\mu_{(2n+2)}}$  matrices  are  obtained by
the
rule
\begin{equation}
\label{GM}
\Gamma^0_{(2n+2)}=\tau^3\otimes I_{(2n)},\quad
\Gamma_{(2n+2)}^{i}=i\tau^2\otimes \sigma^i_{(2n+2)},\quad
i=1,2,\dots,2n+1,
\end{equation}
where $I_{(2n)}$ is a $2^n\times 2^n$ dimensional unit matrix,
$2^n\times 2^n$ dimensional $\sigma^i_{(2n+2)}$ matrices are given
by
\begin{equation}
\label{Sn}
\sigma^1_{(2n+2)}=\Gamma^{D+1}_{(2n)},\quad
\sigma^k_{(2n+2)}=-i\Gamma^{k-1}_{(2n)},\quad k=2,\dots,2n,\quad
\sigma^{2n+1}_{(2n+2)}=\Gamma^0_{(2n)},
\end{equation}
where $\Gamma^{D+1}$--matrix is equal to
\begin{equation}
\label{G5}
\Gamma^{D+1}_{(2n)}=i^{n-1}\Gamma^0_{(2n)}\cdots\Gamma^{2n-1}_{(2n)}
=\tau^1\otimes I_{2n-2}.
\end{equation}
Then the $\Lambda\equiv \Lambda_{(2n)}$ is equal to
\begin{eqnarray}
\label{LM}
&&\Lambda_{(2)}=1,\nonumber\\
&&\Lambda_{(4k+2)}=\underbrace{(I_{(2)}\otimes\tau^2)\times 
\cdots      \times     (I_{(2)}\otimes\tau^2)}_{k\,     {\rm
times}}=(-1)^k\Lambda_{(4k+2)}^T= (-1)^k \Lambda_{(4k+2)}^*=I_{(2)}\otimes
\Lambda_{(4k)}, \nonumber\\
&&\\
&&\Lambda_{(4k)}=\tau^2\otimes\underbrace
{(I_{(2)}\otimes\tau^2)\times
\cdots      \times     (I_{(2)}\otimes\tau^2)}_     {k-1\,{\rm
times}}=(-1)^k \Lambda_{(4k)}^T= 
(-1)^{k} \Lambda_{(4k)}^*=\tau^2\otimes \Lambda_{(4k-2)}\nonumber,
\end{eqnarray}
and 
\begin{equation}
\label{EPS}
\varepsilon\equiv\varepsilon_{(2n)}=(-1)^{n+1}
\end{equation}
Thus  in  any  even dimensional space--time we have a quantum
mechanical  description of the neutral "spin $1/2$" particle
, which is obtained by quantization of the pseudoclassical
theory with the enlarged reparametrization gauge group.

 Lets compare  this  theory  with  the  description of the
neutral (Majorana)  particle  in  the  field  theory.  In the
Dirac representation  the  Majorana  bispinor  $\psi_M$ is
defined by relation
\begin{equation}
\label{MP}
\psi_M=B_{FT} \psi^*_M,\quad B_{FT}=C\Gamma^0,
\end{equation}
where  $C$ is the charge conjugation matrix. As it is known
\cite{JS} in the field theory 
 this  definition  of  the  Majorana spinor is not
contradictory  only  when the dimension of the space--time is
$D~=~2,4~({\rm  mod}8)$.  In  that  case  $B$  is  a
symmetric matrix.

     In  the  quantum  mechanics constructed above the wave
function   $f_M$   satisfies  a  similar  condition  in  any
dimensions:
 \begin{equation}
 \label{QMB}
 f_M=B_{QM}f_M^*,\quad B_{QM}=\left(
\begin{array}{cc}
0&\Lambda^T\\
\Lambda&0
\end{array}
\right) .
 \end{equation}

     To  compare  matrices  $B_{FT}$ and $B_{QM}$ note, that
the   quantum  mechanical  description  corresponds  to  the
Foldy--Wouthuysen  representation  in the field theory. Thus
we  must  first  turn  from  the Dirac representation to the
Foldy--Wouthuysen   representation .
   The  $B_{FT}$  can  be
easily  constructed  explicitly.  Note,  that the inductive
construction  of  $\Gamma $ matrices presented above ensures
the properties of the latter: 
\begin{equation}
\label{PGM}
\Gamma_{(2n)}^{0}=\Gamma_{(2n)}^{0\,T}=\Gamma_{(2n)}^{0+},
\quad\Gamma_{(2n)}^{i}=(-1)^i\Gamma_{(2n)}^
{i\,T}=-\Gamma_{(2n)}^{i\,+},\quad i=1,\dots,2n-1.
\end{equation}
These  allow to explicitly construct the charge conjugation
matrices  $C_{(2n)}$  by  the  rule
\begin{equation}
C_{(2n)}=\beta_{2n}\prod_k\Gamma_{(2n)}^k,
\end{equation}
 where $k$ are even for even $n$ and $k$ are
odd for odd $n$: 
\begin{eqnarray}
\label{CCM}
C_{(8k)}&=&\beta_{8k}i (-1)^k \tau^1 \otimes\Lambda_{(8k)} 
\nonumber\\
C_{(8k+2)}&=&\beta_{8k+2}i (-1)^k \Lambda_{(8k+4)}\nonumber\\ 
C_{(8k+4)}&=&\beta_{8k+4}            (-1)^k           \tau^1
\otimes\Lambda_{(8k+4)} \\
C_{(8k+6)}&=&\beta_{8k+6} (-1)^k \Lambda_{(8(k+1))}\nonumber
\end{eqnarray}
The choice of the $\beta_{2n}$ is a matter of convenience.

Using (\ref{CCM}), the expression for $\Gamma^0$ matrice
and the relations (\ref{LM}), we find
\begin{eqnarray}
\label{FTB}
B_{FT(8k)}&=&\beta_{8k}(-1)^k \tau^2 \otimes\Lambda_{(8k)} 
\nonumber\\
B_{FT(8k+2)}&=&\beta_{8k+2}     (-1)^{k+1}\tau^1     \otimes
\Lambda_{(8k+2)}
\equiv \beta_{8k+2} (-1)^{k+1}B_{QM(8k+2)} \nonumber\\ 
B_{FT(8k+4)}&=&\beta_{8k+4}i        (-1)^{k+1}        \tau^2
\otimes\Lambda_{(8k+4)}
\equiv \beta_{8k+4} (-1)^{k}B_{QM(8k+4)}\\
B_{FT(8k+6)}&=&\beta_{8k+6} i(-1)^k \tau^1 
\otimes\Lambda_{(8k+6)}.\nonumber
\end{eqnarray}
Turning to the Foldy--Wouthuysen representation, the
$B_{FT}$ must be replaced by the matrices
\begin{equation}
\label{FWB}
B_{FT}\rightarrow B_{FW}=U B_{FT}U^T,
\end{equation}
where the unitary matrix $U$ given by
\begin{equation}
\label{FW1}
U=\frac{\hat{\omega}+m+\Gamma^i_{(2n)}\hat{p}{}_i}
{\sqrt{2\hat{\omega}(\hat{\omega}+m)}}
=\frac{1}{\sqrt{2\hat{\omega}(\hat{\omega}+m)}}
\left(
\begin{array}{cc}
\label{FWT}
(\hat{\omega}+m)&\sigma^i_{(2n)} p_i\\
-\sigma^i_{(2n)} p_i &(\hat{\omega}+m)
\end{array}
\right)
\end{equation}
 connects  the Dirac and Foldy-- Wouthuysen representations.
Using   the  explicit  expressions  for  $B_{FT}$  and  the
properties (\ref{LM}) of $\Lambda$ matrices we obtain:
\begin{equation}
\label{FWQM1}
B_{FW}\equiv B_{FT}.
\end{equation}
Using  (\ref{FTB}) we see, that in the space--time dimensions
where  the  Majorana particle exists, namely when $D=2,4({\rm
mod}8)$,  the following relation holds (with appropriate choice
of $\beta$) :
\begin{equation}
\label{FWQM}
B_{FW}=B_{QM},
\end{equation}
and hence the quantum mechanical description of the neutral
particle in the space--time dimensions $D=2,4({\rm mod}8)$ coincides
with the field theoretical description of the Majorana particle
in the Foldy-- Wouthuysen representations.

Note again, that in quantum mechanics  the neutral particle
exists in any even dimensions $D=2n$.

\section{$D=2n$--dimensional Weyl particle}
\indent

Consider a theory with the action given by the expression
\begin{eqnarray}
\label{ACTMW}
S&=&\frac{1}{2}\int
d\tau\left[\frac{1}{e}{\left(z^{\mu}\right)}^2+e m^2-
i\left(\xi_\mu\dot{\xi}^\mu -\xi_{D+1}\dot{\xi}_{D+1}\right)
+im\chi\xi_{D+1}\right],\\
z^\mu&=&\dot{x}^\mu -
\frac{i}{2}\chi\xi^\mu -
\frac{(-i)^{\frac{D-2}{2}}}{(D-2)!}
\varepsilon^{\mu\nu\lambda_1\dots \lambda_{D-2} }
b_\nu\xi_{\lambda_1}\cdots \xi_{\lambda_{D-2}}+\tilde{\alpha} b^\mu
\nonumber
\end{eqnarray}
Here  $b^\mu$  are  additional fields, $\tilde{\alpha}$ is a
constant ; $b^\mu, \,\tilde{\alpha}$  are  Grassmann  even.

This theory is gauge invariant under the transformations:
 the  re\-pa\-ra\-me\-tri\-zation transformations
\begin{eqnarray}
\label{RT}
\delta x^\mu&=&u\dot{x}^\mu,  \quad\delta
e=\frac{d}{d\tau}(ue),  \quad\delta b^\mu=
\frac{d}{d\tau}(ub^\mu),\nonumber\\
\delta\xi^\mu&=&u\dot{\xi}^\mu,  \quad
\delta\chi=\frac{d}{d\tau}(u\chi),\quad
\delta\xi_{D+1}=u\dot{\xi}_{D+1},
\end{eqnarray}
with    the    even    parameter    $u(\tau)$,    supergauge
transformations
\begin{eqnarray}
\label{ST}
\delta x^\mu&=&iv\xi^\mu,\quad\,\delta e=iv\chi,\quad\,\delta
b^\mu=0,\,\quad
\delta\xi^\mu=v\frac{z^\mu}{e},\quad\,\delta \chi=2 \dot{v},\quad
\delta\xi_{D+1}=vm,\nonumber\\
z^\mu&=& \dot{x}^\mu -
\frac{i}{2}\chi\xi^\mu -
\frac{(-i)^{\frac{D-2}{2}}}{(D-2)!}
\varepsilon^{\mu\nu\lambda_1\dots \lambda_{D-2} }
b_\nu\xi_{\lambda_1}\cdots \xi_{\lambda_{D-2}}+\tilde{\alpha} b^\mu
\end{eqnarray}
with   the   odd   parameter   $v(\tau)$.

Let us find the equations of motion, corresponding to the
variations of the action over $e$ and $b_\mu$:
\begin{eqnarray}
\label{EM}
&&p_\mu^2-m^2=0,\quad \quad p_\mu\equiv\frac{1}{e}z_\mu,\nonumber\\
&&\frac{(-i)^{\frac{D-2}{2}}}{(D-2)!}
\varepsilon^{\mu\nu\lambda_1\dots \lambda_{D-2} }
p_\nu\xi_{\lambda_1}\cdots \xi_{\lambda_{D-2}}+\tilde{\alpha}
p^\mu=0.
\end{eqnarray}
Multiplying the second of these equations by $p_\mu$ we find the
condition
\begin{equation}
\tilde{\alpha} m^2=0.
\end{equation}
Thus the theory with the action (\ref{ACTMW}) is not
contradictory only for the choice of the parameters
$\tilde{\alpha}=0$ or (when $\tilde{\alpha}\neq0$) $m=0$. In what
follows we will consider the case of $\tilde{\alpha}\neq0$, hence
we must put $m=0$ in  (\ref{ACTMW}). Thus consider a theory with
the action given by the expression
\begin{eqnarray}
\label{ACTW}
S=\int d\tau\left[\frac{1}{2e}\left(\dot{x}^\mu -
\frac{i}{2}\chi\xi^\mu -
\frac{(-i)^{\frac{D-2}{2}}}{(D-2)!}
\varepsilon^{\mu\nu\lambda_1\dots \lambda_{D-2} }
b_\nu\xi_{\lambda_1}\cdots \xi_{\lambda_{D-2}}+\tilde{\alpha} b^\mu
\right)^2-
\frac{i}{2}\xi_\mu\dot{\xi}^\mu \right],
\end{eqnarray}
which is a generalization to the space--time dimension $D=2n$
of  the  pseudoclassical theory of Weyl particle \cite{GGT} (it
turns out, that the variable $\xi_{D+1}$ can be omitted from the
action (\ref{ACTMW}) at $m=0$).
Apart from the invariance under the transformations (\ref{RT})
and (\ref{ST}) the  action  (\ref{ACTW})  is  invariant under the
additional gauge   transformations   \cite{GGT}
\begin{eqnarray}
\label{WT}
\delta x^\mu&=&\frac{(-i)^{\frac{D-2}{2}}}{(D-2)!}
\varepsilon^{\mu\nu\lambda_1\dots \lambda_{D-2} }\eta_\nu
\xi_{\lambda_1}\cdots \xi_{\lambda_{D-2}}-\tilde{\alpha}
\eta^\mu,\nonumber\\
\delta \xi^\mu&=&\frac{1}{e}\frac{(-i)^{\frac{D}{2}}}{(D-3)!}
\varepsilon^{\mu\nu\delta\lambda_2\dots \lambda_{D-2} }
\eta_\nu z_\delta\xi_{\lambda_2}\cdots
\xi_{\lambda_{D-2}},\\
\delta b^\mu&=&\frac{d}{d\tau}(\eta^\mu),  \quad\delta\chi=-2\eta_\nu(p^\nu
\xi^\sigma-p^\sigma \xi^\nu)b_\sigma \delta_{D4},  \quad \delta e=
-2i\eta_\nu \xi^\nu\xi^\sigma b_\sigma \delta_{D4},\nonumber
\end{eqnarray}
with the even parameter $\eta_\nu(\tau)$.

Acting in the standard way we obtain the
  canonical  hamiltonian  of  the  theory, which is given by the
expression
\begin{eqnarray}
\label{CH}
H &=&\dot{x}^\mu p_\mu+\dot{\xi}^\mu\pi_\mu
-L=\nonumber\\
&=&\frac{e}{2}p^2+\frac{i}{2}\chi p_\mu \xi^\mu-
\left(\frac{(-i)^{\frac{D-2}{2}}}{(D-2)!}
\varepsilon^{\nu\mu\lambda_1\dots \lambda_{D-2} }
p_\mu\xi_{\lambda_1}\cdots \xi_{\lambda_{D-2}}+\tilde{\alpha}
p^\nu\right)b_\nu,
\end{eqnarray}
primary constraints
\begin{equation}
\label{CNST}
\Phi_1^{(1)}=\pi_e, \quad \,  \Phi_2^{(1)}=\pi_\chi, \,\quad
\Phi_{3\mu}^{(1)}=\pi_\mu-\frac{i}{2}\xi_\mu,
\quad\,\Phi_{4\mu}^{(1)}=\pi_\mu^b
\end{equation}
and the secondary constraints
\begin{eqnarray}
\label{SC}
\Phi_1^{(2)}&=&|p_0|-\omega,  \quad
\Phi_{2}^{(2)}=p_\mu\xi^\mu,\nonumber\\
\Phi_{3\mu}^{(2)}&\equiv&T_\mu=\frac{(-i)^{\frac{D-2}{2}}}{(D-2)!}
\varepsilon_{\mu\nu\lambda_1\dots \lambda_{D-2} }
p^\nu\xi^{\lambda_1}\cdots \xi^{\lambda_{D-2}}+\tilde{\alpha} p_\mu,
\end{eqnarray}
 where $\omega=|\vec{p}|,\,\vec{p}=(p_k),\,k=1,\dots,D-1$.
One can see now, that the canonical hamiltonian $H$ is
equal  to zero  on  the  constraints  surface, as it was
expected to be.

The  constraints $F\equiv (\Phi^{(1)}_1,  \,  \Phi^{(1)}_2,  \,
\Phi^{(1)}_{4\mu},  \,  \Phi^{(2)}_1),  \,  \Phi^{(2)}_{3\mu}$
 are  first class. Apart from them there is
one more  first class constraint $\varphi$, which is a
linear combination of the constraints
$\Phi_{3\mu}^{(1)}$,  $\Phi^{(2)}_2$:
\begin{equation}
\label{FCC}
\varphi=p^\mu \Phi_{3\mu}^{(1)}+i\Phi_2^{(2)}=p_\mu
\pi^\mu+\frac{i}{2}p_\mu \xi^\mu.
\end{equation}
Adhering  to  the  quantization  method, when already at the
classical  level  all  gauge  degrees  of  freedom are fixed
\cite{GT2}  ,  we  must introduce into the theory additional
constraints, 
 equal  in  number  to  that of all first class
constraints and conjugated to the latter. 
However, as it was
noted in \cite{GGT}, the constraints $\Phi_{3\mu}^{(2)}$ for
$D\geq 4$ are at least quadratic functions of the variables
 $\xi^\lambda$,   thus   complicating  the  introduction  of
additional constraints conjugated to $\Phi_{3\mu}^{(2)}$.
 For  this  reason  the constraints $\Phi_{3\mu}^{(2)}$ after
quantization  will be used as conditions on the state vectors.
For  the remaining first class constraints $F$ and $\varphi$
 we will introduce additional constraints $\Phi^G$ in the form
\begin{equation}
\label{GF}
\Phi^G_{1}=x^0-\kappa \tau, \quad \,  \Phi^G_{2}=\chi, \quad \,
\Phi^G_{3}=e-\frac{\kappa}{p_0},\quad\, \Phi^G_{4\nu}=b_\nu, \quad \,
\Phi^G_{5}=\xi^0.
\end{equation}
 To go over to time-independent
set  of  constraints  we  perform a canonical transformation
(\ref{CTR}),  after  which  the hamiltonian of the system on
the constraint surface is given by the expression
\begin{equation}
\label{HAM}
H=\omega=|\vec{p}|.
\end{equation}
The quantization of the theory is carried out using the formulae
of the previous section with $m=0$.

 Note
,  that  in  comparison with the previous section, the system  of
second-class constraints in this theory contains new constraints
$\Phi_{4\mu}^{(1)}=\pi_\mu^b\approx  0,$ $\,\Phi^G_{4\mu}=
b_\mu\approx   0$.   However  they  have  a special  form
\cite{GTY} and  they  do  not  affect  the  final Dirac brackets
(the  variables  $b_\mu$  and  $\pi_\mu^b$  can be excluded from
the theory using the constraints).

Now    using    the    expressions    (\ref{QOR})   for
$\hat{\xi}^i$ with $m=0$,  we can find the expressions for the operators
$\hat{T}_\mu$,   which   correspond   to   the  first  class
constraints $\Phi_{3\mu}^{(2)}$:
\begin{equation}
\label{TO}
\hat{T}_{\mu}=\left(\frac{\hbar}{2}\right)^{(D-2)/2}\hat{p}_\mu \hat{T},
\quad
\hat{T}=\hat{\kappa}\frac{\hat{p}^i\Sigma^i}{\hat{\omega}}
-\alpha,  \quad\hat{p}_0=-\hat{\kappa}\hat{\omega},
\quad
\alpha=\left(\frac{\hbar}{2}\right)^{-(D-2)/2}\tilde{\alpha}.
\end{equation}
To deduce these relations we used the equality
$\varepsilon_{01\dots(D-1)}=-\varepsilon_{12\dots(D-1)}=-1$,
and also the relation
\begin{equation}
\frac{(-i)^{\frac{D-2}{2}}}{(D-2)!}
\varepsilon^i_{\,\,j_1\cdots j_{D-2} }
\sigma^{j_1}\cdots \sigma^{j_{D-2}}=\sigma^i
\end{equation}
from the $\sigma^i$--matrix algebra in $(D-1)$--dimensional space
\cite{CA}. The $\alpha$ is the eigenvalue of the chirality
operator $\hat{\vec{p}}\vec{\sigma}/\hat{\omega}$ and in
quantum mechanics can have the values $+1$ or $-1$.

As   it   was   already   mentioned   above,  the  operators
$\hat{T}_\mu$   will  be  used to impose conditions on the
physical  state  vectors. For the $D$--dimensional space--time
the  state  vector  has in  general
$2^{D/2}$ components (in   massless   case  this  representation  is
reducible). Representing the state vector $f$ in the form
 \begin{equation}
\label{SV}
f=\left(
\begin{array}{c}
f^1\\f^2
\end{array}
\right),
\end{equation}
where    $f^1$   and   $f^2$   are   $2^{(D-2)/2}$
dimensional columns,
we  write  down  the  equations  for the state vector in the
form:
\begin{equation}
\label{TO2}
\hat{T} f=0.
\end{equation}

It is natural to interpret the quantum mechanics constructed
above as a theory of Weyl particle
in the Foldy--Wouthuysen representation.
Indeed, using the realization (\ref{KAP}) for $\hat{\kappa}$ we
write the operator $\hat{T}$ in the form
\begin{equation}
\label{TO4}
\hat{T}\equiv \hat{T}_{FW}=\Gamma^0_{(2n)}
\frac{\hat{p}^i\Sigma^i}{\hat{\omega}}-\alpha
\end{equation}

Consider also the Shr\"odinger equation
$(i\partial/\partial\tau-\hat{H})  f=0$, which describes the
evolution  of the state-vector $f$ with respect to parameter
$\tau$ . Being rewritten in terms of
the physical time $x^0=\kappa\tau$
it takes the form
\begin{equation}
\label{SE}
(i\partial/\partial x^0-\Gamma^0_{(2n)} \hat{\omega})f=0.
\end{equation}
Applying  the  unitary  Foldy--Wouthuysen  transformation for
$m=0$
in  $D$  dimensional  space--time
\begin{equation}
\label{FW}
 f=U \psi,  \quad
U=\frac{\hat{\omega}+\Gamma^i_{(2n)}\hat{p}^i}{\hat{\omega}\sqrt{2}},
\end{equation}
 where $\psi$ is the wave function in the Dirac representation,
 we find \cite{GT2,GG1} that the  Shr\"odinger equation
transforms into Dirac equation, the expressions for the Lorentz
generators $\hat{J}^{\mu\nu}$ transform into standard expressions
for the Lorentz generators in the Dirac representation.
Furthermore, by direct calculation one can prove that the
operator $\hat{T}_{FW}$ transforms into the $\hat{T}_D$ operator
\begin{equation}
\label{TDO}
\hat{T}_D=U^+\hat{T}_{FW} U=\Gamma_{(2n)}^{D+1} - \alpha,
\end{equation}
which is proportional to a standard Weyl projector in the Dirac
representation.

Thus we see that the quantum mechanical description constructed
here after the Foldy--Wouthuysen transformation  turns into
the Dirac description of the Weyl particle. Hence the  quantum
mechanics constructed above describes the Weyl particle in
the Foldy--Wouthuysen representation.

\section{ Quantum mechanics of Majorana--Weyl particle}
\indent

As it was discussed in section 2, the invariance of the
action under the reparametrization transformation
$\tau\rightarrow-\tau$ allows, after the introduction of this
transformation in the gauge group,
 to describe a neutral particle in the QM. We will apply
this ideology to the action (\ref{ACTW}) of the previous section.

For the invariance of the action with respect to 
the transformation  $\tau\rightarrow-\tau$
 apart from (\ref{TI}) we need a transformation
\begin{equation}
\label{TIB}
b^\mu(\tau)\rightarrow -b^\mu(-\tau).
\end{equation}

Now the action (\ref{ACTW})  will be invariant
 under the transformations (\ref{TI}), (\ref{TIB}) if
\begin{equation}
\frac{D-2}{2}=2k,
\end{equation}
i.e. in the space--time dimensions $D=4k+2$. In these dimensions
we will quantize the action (\ref{ACTW}) with the extended gauge
group and will compare this result with the description of
Majorana--Weyl particle in the field theory.
 Adhering to the first approach to quantization, when instead of
 the gauge $\Phi=x^0-\kappa\tau$ we choose the
 $\Phi^\prime=x^0-\tau$ gauge, we simply restrict the physical
 states space to the sector $\kappa=+1$ of the Weyl particle in QM.
 The wave function is described now by a $2^{(D-2)/2}$ column
 \begin{equation}
\label{MWWF}
f=f^1,
\end{equation}
while the Weyl  condition has the form
\begin{equation}
\label{WC1}
\hat{P}_\alpha f^1\equiv\frac{1}{2}\left(\frac{\hat{\vec{p}}\vec{\sigma}}
{\hat{\omega}}-\alpha\right)f^1=0,
\end{equation}
where the operators $\hat{P}_\alpha$ is defined by relation
\begin{equation}
\hat{P}_\alpha=\frac{1}{2}\hat{T}|_{\kappa=+1}
\end{equation}

For the comparison with the field theory it is convenient to
introduce the "bispinor" $f_M$
 \begin{equation}
\label{MWBS}
f_M=\left(
\begin{array}{c}
f^1\\\Lambda f^{1*}.
\end{array}
\right).
\end{equation}

The same wave function turns out in the second approach to
quantization, when we factorize the total states space of the
Weyl particle with respect to the gauge symmetry $\tau\rightarrow
-\tau$. Only now one more condition must be fulfilled
\begin{equation}
\hat{P}_{-\alpha}\Lambda f^{1*}=0,
\end{equation}
from which follows that the relation
\begin{equation}
\hat{P}_{-\alpha}\sim\Lambda \hat{P}_\alpha^* \Lambda^{-1}
\end{equation}
must hold.
Using   the   explicit   formulae  (\ref{SIG}), (\ref{EPS})  and
the properties (\ref{LM}) of $\Lambda$--matrices we convince
ourselves, that this relation is true only for dimensions
$D=4k+2$. Thus two approaches to quantization give equivalent
answers: the quantum mechanics in space--time dimensions $D=4k+2$
with enlarged gauge group describes a neutral  (the antiparticle
coincides with the particle) Weyl particle. Since this
description can be equivalent to  the field theoretical
description of the Majorana particle only when $D=8k+2$ and
$D=8k+4$, we find, that in the dimensions $D=8k+2$ the
constructed QM is equivalent to field theoretical description of
Majorana--Weyl particle in the Foldy--Wouthuysen representation.
The singling out of the  dimensions $D=8k+2$ is in agreement with
the fact, that in the field theory Majorana--Weyl particle exists
only in these dimensions \cite{JS} (in QM the neutral Weyl particle
exists also in dimensions $D=8k+6$).

This research was partially supported  by the grants
RFBR/INTAS-96-829 and 211-5291 YPI of
the German Bundesministerium f\"ur Forschung und Technologie.
I.V.Tyutin was supported in part by grant \# M21300 from
international Science Foundation and Government of the Russian
Federation and by European Community Commission under contract
INTAS-94-2317.


\end{document}